\documentclass[a4paper,12pt]{article}  
\usepackage[dvips]{graphicx}
\usepackage{amssymb}
\usepackage{cite}
\newcommand{\goes}{\rightarrow} 
 
\newcommand{\GeV}{\; \mathrm{GeV}} 
\newcommand{\TeV}{\; \mathrm{TeV}} 
\newcommand{\lapproxeq}{\lower .7ex\hbox{$\;\stackrel{\textstyle  
<}{\sim}\;$}} 
\newcommand{\gapproxeq}{\lower .7ex\hbox{$\;\stackrel{\textstyle  
>}{\sim}\;$}} 
\newcommand{\stackdown}[2]{\lower 1.4ex\hbox{$\;\stackrel{\textstyle{#1}}  
{\scriptstyle{#2}}\;$}}
\newcommand{\beq}{\begin{equation}} 
\newcommand{\eeq}{\end{equation}} 
\newcommand{\bea}{\begin{eqnarray}} 
\newcommand{\eea}{\end{eqnarray}}
\newcommand{\lsp}{\tilde{\chi}}
\newcommand{\mlsp}{m_{\lsp}}

\newcommand{\relic}{\Omega_{\lsp}\,h_0^2} 
\newcommand{\sxsec}{\sigma_{scalar}} 
\newcommand{\almuon}{\alpha_{\mu}^{\mathrm{SUSY}}}
 
\newcommand{\etal}{\textit{et. al.}}

\makeatletter 
\def\slash{\@ifnextchar[{\fmsl@sh}{\fmsl@sh[0mu]}} 
\def\fmsl@sh[#1]#2{%
  \mathchoice 
    {\@fmsl@sh\displaystyle{#1}{#2}}%
    {\@fmsl@sh\textstyle{#1}{#2}}%
    {\@fmsl@sh\scriptstyle{#1}{#2}}%
    {\@fmsl@sh\scriptscriptstyle{#1}{#2}}} 
\def\@fmsl@sh#1#2#3{\m@th\ooalign{$\hfil#1\mkern#2/\hfil$\crcr$#1#3$}} 
\makeatother 

\begin{document} 
\begin{titlepage} 
 
\begin{flushright} 
\parbox{6.9cm}{ ACT-07/01, CTP-TAMU-24/01\\
                CERN-TH/2001-168, UA/NPPS-8-01\\
                HEPHY-PUB 741/01\\
                hep-ph/0107151 } 
\end{flushright} 
\begin{centering} 
\vspace*{1.5cm} 
 
{\large{\textbf {Dark Matter Direct Searches and the 
Anomalous Magnetic Moment of Muon
}}}\\
\vspace{1.4cm}

{\bf A.~B.\ Lahanas} $^{1}$, \, 
{\bf D.~V.~Nanopoulos} $^{2}$  \, and \, {\bf V.~C.~Spanos} $^{3 }$  \\ 
\vspace{.8cm} 
$^{1}$ {\it University of Athens, Physics Department,  
Nuclear and Particle Physics Section,\\  
GR--15771  Athens, Greece}\\ 

\vspace{.5cm} 
$^{2}$ {\it Department of Physics,  
         Texas A \& M University, College Station,  
         TX~77843-4242, USA, 
         Astroparticle Physics Group, Houston 
         Advanced Research Center (HARC), Mitchell Campus, 
         Woodlands, TX 77381, USA, and \\ 
         Academy of Athens,  
         Chair of Theoretical Physics,  
         Division of Natural Sciences, 28~Panepistimiou Avenue,  
         Athens 10679, Greece}  \\ 
 
\vspace{.5cm} 
$^{3}$ {\it  Institut f\"ur Hochenergiephysik der \"Osterreichischen Akademie
der Wissenschaften,\\
A--1050 Vienna, Austria }  \\ 
\end{centering} 
\vspace{1.8cm} 
\begin{abstract}

In the framework of 
the Constrained Minimal Supersymmetric
Standard Model (CMSSM) we discuss the impact of  the recent experimental
information, especially from E821 Brookhaven experiment  
on  $g_{\mu}-2$ along with the light Higgs boson mass bound
from LEP, to the Dark Matter direct searches.
Imposing these experimental bounds, the
maximum value of the spin-independent neutralino-nucleon
cross section  turns out to be of the order of $10^{-8}$ pb
for large values of $\tan\beta$ and low $M_{1/2}, m_0$.
The effect of the recent experimental bounds
is to decrease the maximum value of the cross section
by about an order of magnitude, demanding the
analogous sensitivity from the direct Dark Matter detection
experiments.

\end{abstract} 
\end{titlepage} 
\newpage 
\baselineskip=20pt 

Supersymmetry, or fermion-boson symmetry, is an omnipotent and ubiquitous
element in our efforts to construct a unified theory of all
fundamental interactions observed in nature. At very high energies, close
to the Planck scale ($M_P$) it is indispensable  in 
constructing consistent string
theories, thus dubbed superstrings. At low energies ($\sim 1 \TeV$)
it seems unavoidable if the gauge hierarchy problem is to
be resolved. Such a resolution provides a measure of the supersymmetry
breaking scale $M_{SUSY} \thickapprox \mathcal{O}(1 \TeV) $. 
There is, albeit circumstantial or indirect, evidence for such a
 low-energy supersymmetry breaking scale, from the unification
of the gauge couplings \cite{Kelley} and from the apparent lightness
of the Higgs boson as determined from precise electroweak measurements,
mainly at LEP \cite{EW}. Furthermore, such a low energy SUSY breaking
scale is also favored cosmologically. As is well known, $R$-parity
conserving  SUSY models, contain in the sparticle spectrum a stable,
neutral particle, identifiable with the lightest neutralino ($\lsp$),
referred as the LSP \cite{Hagelin}. 
One can then readily show \cite{Hagelin} that
such a LSP with mass, as low-energy SUSY entails, in the
 $ 100\GeV - 1\TeV $ region, may indeed provide the right form and amount
of the highly desirable astrophysically and cosmologically Dark Matter (DM).
As times goes by, the experimental evidence for DM, from
different quarters, strengthens in such a way, that it has assumed a central
role in the modern cosmology. The most recent evidence, coming from
the observation of the first three acoustic peaks in the Cosmic
Microwave Background (CMB) radiation small 
angle ($\theta \lesssim \mathcal{O}(1^0)$ anisotropies \cite{cmb}, is
of  tantalizing importance.
It is not only provides  strong support to a flat ($k=0$ or $\Omega_0=1$),
inflationary Universe, but it also gives an unprecedented determination of
$\Omega_M h_0^2 \thickapprox 0.15 \pm 0.05$,
which taking into account the simultaneously determined baryon density
$\Omega_B h_0^2 \thickapprox 0.02$, and the rather minute 
neutrino density suggests
\beq
\Omega_{DM} h_0^2 = 0.13 \pm 0.05
\label{cbound}
\eeq
One then is tempted to combine this recently determined DM density,
assuming, as we do here, that it is all due to neutralini
(i.e. $\Omega_{DM} \equiv \Omega_{\lsp}$), with other
presently available constraints from particle physics, in order
to find out what is the chances of observing, soon or in the
near future, DM directly in the laboratory by elastic 
neutralino-nucleus scattering, from the energy deposition in the
detectors \cite{Goodman}.
These particle physics constraints include the lower bound on the
mass of the Higgs bosons ($m_{h} \geq 113.5 \GeV$) provided by LEP \cite{LEP},
the allowed region for $b \goes s \gamma$, at $95\% $ CL range
($2.33 \times 10^{-4} < \mathcal{B}(b \goes s \gamma)
                            <4.45 \times 10^{-4}$) \cite{Alam},
and the recent results from the BNL E821 experiment \cite{E821} on the 
anomalous magnetic moment of the muon 
($\delta \alpha_{\mu} =43 (16)\times 10^{-10}$), assuming, as we do here,
that is all due (at the 1 or 2 $\sigma$ level) to low-energy
supersymmetry. 
It should be stressed that the possibility of a rather sizeable positive
contribution to $g_{\mu}-2$ from low energy SUSY in the region of
large $\tan\beta$ and $\mu > 0$ where the $b \goes s \gamma$ constraint
weakens considerably, has been long strongly emphasized \cite{LNW}.
It is amusing to notice, that in our previous analysis of the direct
DM searches \cite{LNSd}, done {\em before} the the BNL E821
announcement, we had paid particular interest in the large $\tan\beta$
region, since it provided the higher possible rates for
direct DM detection! 
Similar results are presented in Ref.~\cite{kim}.
Actually, as we have 
stressed for some time now \cite{LNS,LNSd}, one way to get the ``right''
amount of the neutralino mass density ($\relic$), even for relative
high values of $m_0$ and $M_{1/2}$, is to move to the large 
$\tan\beta$ region, because efficient neutralino annihilation
directly through $A$ and $H$ poles, occurs.
The annihilation cross sections increase
with $\tan\beta$: couplings $A \lsp \lsp$ and $A\tau \bar{\tau}$, 
$A b \bar{b}$ increase, while $m_A$ decreases, thus one may also
expect a rather appreciable increase in the elastic $\lsp$-nucleon
cross section, as is indeed the case.
It may turn out, if the BNL E821 result is due to low energy SUSY,
that the imposed lower bounds on $m_0$, $M_{1/2}$ and lower bounds on
$\tan\beta$ \cite{ENO,g-2,Rosz,LS} make the direct neutralino annihilation,
through the $A$, $H$ poles, the major mechanism for getting the 
right amount of DM \cite{LNSd,LNS,Drees}, 
as well as as being consistent with all
available constraints \cite{ENO,LS}.

Before presenting our results  
we shall give a brief account on the numerical analysis employed in this
paper. This will be useful in comparing our results with those of other
authors. 
In our analysis we use two-loop renormalization group equations (RGE), in
the $\overline{DR}$ scheme, for all masses and couplings involved, defining 
the unification scale $M_{GUT}$ as the point at which the gauge
couplings ${\hat \alpha}_1$ and ${\hat \alpha}_2$ meet.
We do not enforce unification of ${\hat \alpha}_3$ gauge coupling with  
with ${\hat \alpha}_{1,2}$ at $M_{GUT}$.
The experimental value of the $\overline{MS}$
strong coupling constant at $M_Z$, which we consider as input,  
is related to ${\hat \alpha}_3$ through
$\alpha_s(M_Z)={\hat \alpha}_3(M_Z) / (1 - \Delta {\hat \alpha}_3 )$.  
$\Delta {\hat \alpha}_3$ represent the threshold corrections which 
affect significantly the value of ${\hat \alpha}_3$ at $M_Z$ and hence, 
through RGE, its value at $M_{GUT}$. The latter turns out to be different
from ${\hat \alpha}_{1,2}(M_{GUT})$, reflecting the fact that  
gauge coupling unification is impossible to implement in the constrained
scenario with universal boundary conditions for the soft masses. 
For the determination of the
gauge couplings ${\hat \alpha}_{1,2}$ we use as inputs the
electromagnetic coupling constant $a_0$ 
the value of the Fermi coupling constant $G_F$,
and the $Z$-boson mass $M_Z$. From these we determine the
weak mixing angle, through 
${\hat s}^2 {\hat c}^2 ={\pi \alpha_0}/{\sqrt{2} M_Z^2 G_F}
{(1-\Delta {\hat r}})$, and  
the value of the electromagnetic coupling constant at $M_Z$.
With $\Delta {\hat r}$ we denote the  correction to
the muon decay amplitude.
In the
$\overline{DR}$ scheme the latter is related to $a_0$  through
${\hat \alpha}(M_Z) = a_0 /(1 - \Delta {\hat \alpha}_{em})$, where
$\Delta {\hat \alpha}_{em}$ are the appropriate threshold
corrections (see Ref.~\cite{BMPZ}). The input value of the strong coupling
constant is taken within the experimental range
$\alpha_s(M_Z)=0.1185 \pm 0.002$. 

In running the RGE's, as arbitrary parameters we take  
are as usual the soft SUSY breaking parameters $m_0, M_{1/2}, A_0$ the
value of 
$\tan \beta$ and the sign of the Higgsino mixing parameter $\mu$.
The top and tau physical masses,  
$M_t, M_{\tau}$, as well as the $\; \overline{MS}$ bottom
running mass $m_b(m_b)$ are also inputs.
As default values we consider 
$M_t = 175 \GeV, M_{\tau} = 1.777\GeV$ and $m_b(m_b) = 4.25\GeV$ although we
allow for variations within their experimentally allowed region.  
The determination of the bottom and tau running masses at $M_Z$ is done  
by running $S{U_c}(3) \times U_{em}(1)$ $\; \overline{MS}$ RGE's,  
using three-loop RGE's for the strong coupling constant. We also include 
two-loop QED corrections, as well as two-loop contributions from the
interference of the QCD and QED corrections. 
The running $ \overline{MS}$ masses at $M_Z$ are then converted to 
$ \overline{DR}$ in the usual way. From these we can extract 
the corresponding Yukawa couplings at $M_Z$.
We point out that the important QCD as well as
the supersymmetric gluino, sbottom and chargino, stop  corrections to 
the bottom mass are duly taken into account.
For the determination of the top Yukawa coupling at $M_t$ we relate its 
pole and running masses taking into account all dominant radiative
corrections. By running the RGE's we can have the value of the top Yukawa
coupling at $M_Z$. 

The determination of the Higgs and Higgsino mixing parameters, 
$m_3^2$ and  $\mu$, is a more subtle issue. These are obtained by solving 
the minimization conditions with the one-loop corrected 
effective potential with all particle contributions taken into account. 
Since large values of $\tan \beta$ cause large logarithmic corrections,    
invalidating perturbation expansion, we solve
the minimization equations taking as reference scale the average stop scale
$Q_{\tilde t}\simeq\sqrt{m_{{\tilde t}_1} m_{{\tilde t}_2} }$. At this scale
the corrections are numerically small and hence perturbatively valid. 
Thus in each
run we determine $m_3^2(Q_{\tilde t}), \mu(Q_{\tilde t})$.  
The values of
$m_3^2(M_Z), \mu(M_Z)$, whenever needed, can be found by solving the RGE's
having as initial conditions the values of these quantities at $Q_{\tilde t}$.

For the calculation of the lightest supersymmetric particle (LSP) 
relic abundance, we solve the Boltzmann equation  
numerically using the machinery
outlined in Ref.~\cite{LNS}. In this calculation the coannihilation effects, 
in regions where $\tau_R$ approaches in mass the LSP, which is a high
purity Bino, are properly taken into account. 

Before embarking to analyse our numerical findings 
it would be beneficial to
review the physical mechanism through which the scalar, i.e.
spin-independent,
$\lsp$-nucleon cross section ($\sxsec$) is enhanced, to levels
approaching the sensitivity of ongoing experiments.
The $\sxsec$ is enhanced in the region of the parameter
space where $\tan\beta$ is large \cite{LNSd}
\footnote{Enhancement of $\sxsec$ is also
possible  in the context of the so-called
focus point supersymmetry scenario \cite{focus}, 
where $m_0 > 1.5\TeV$, yet such large
values of $m_0$ are not favourable by the recent $g_{\mu}-2$ data.}. 
The dominant contribution to this regime is the Higgs boson  exchange.
For given inputs $m_0, M_{1/2}, A_0$  and the sign of $\mu$, 
Higgs masses decrease as $\tan \beta$ increases.
Hence the contribution of Higgs bosons to 
neutralino--quark elastic cross section becomes
more important in the large $\tan \beta$ regime. Such a decrease in the mass
is not sufficient by itself to increase $\sxsec$.
The major role in this
increase plays the coupling of 
the $CP$-even heavy Higgs whose coupling to $d$-quark is proportional to
$\frac{\cos \alpha}{\cos \beta}$, which is proportional to
$\tan \beta$, when the latter becomes large.
The coupling of the light $CP$-even Higgs, unlike
the heavy Higgs case, does not grow with increasing
${\tan \beta}$ but stays constant of order unity.
Therefore despite the fact that the heavy $CP$-even Higgs is heavier 
than its light $CP$-even counterpart, its contribution
may be much larger in the large $\tan \beta $  region,  
due to its enhanced coupling to $d$-quark. 
In the
large $\tan \beta$ regime the mass of the heavy $CP$-even Higgs
can be approximated \cite{LNSd} by the relation
\beq
m_H^2 \approx  m_h^2 +  m_A^2 - m_Z^2 - \epsilon  \,,
\label{heavy}
\eeq
where $\epsilon$ represents the leading stop 1-loop corrections to the
$CP$-even Higgs masses.
From this is obvious that the lowest $m_H$ 
values are obtained in the region where $m_A$ is
light and $m_h$ is close to its lower experimental bound.
Moreover it  is worth pointing out that in the large $\tan \beta$ region 
neutralino relic densities decrease as we have already emphasized
due to both the decrease of the pseudoscalar mass, whose
exchange in $\lsp \lsp \rightarrow b {\bar b} \;,\;\tau {\bar \tau}$
processes is less suppressive, and the
increase of the $\lsp \lsp A$ as well as the
$A b \bar{b}$ and  $A \tau \bar{\tau}$ couplings. 
The smallness of the LSP's 
Higgsino component is compensated by the largeness of $\tan \beta$ yielding
neutralino annihilation cross sections compatible with the recent
astrophysical data.
Hence there are regions in which we can obtain both low relic densities
and high $\sigma_{scalar}$.

\begin{figure}[t] 

\centering\includegraphics[height=9cm,width=10cm]{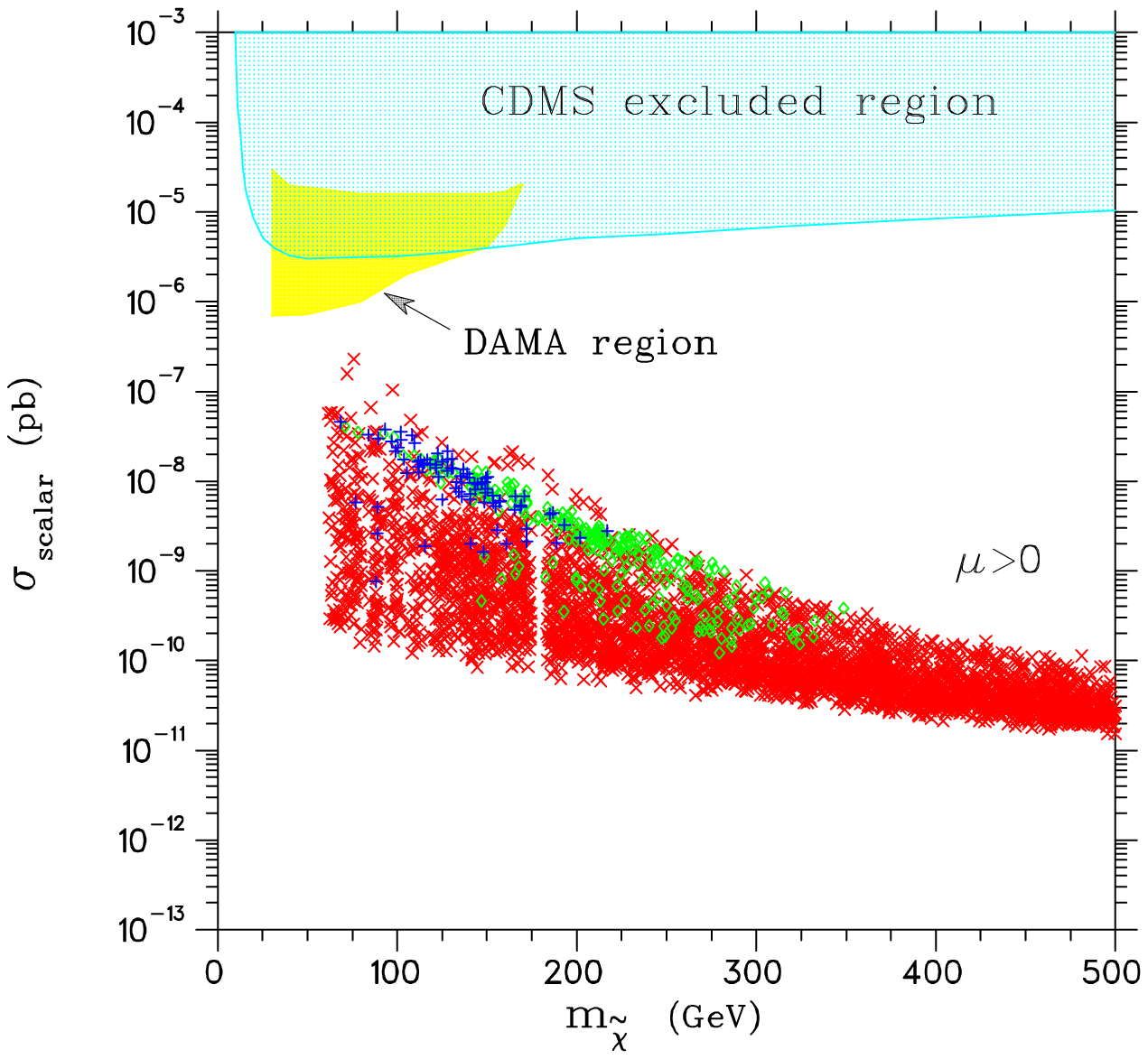}
\caption[]{
Scatter plot of
the scalar neutralino-nucleon cross section
 versus $\mlsp$, from a
random sample of 45,000 points.
On the top of the figure the CDMS excluded region and
the DAMA sensitivity region are illustrated.  
Pluses ($+$) are points within the E821 experimental region
$\almuon = ( 43.0 \pm 16.0 ) \times 10^{-10}$ and
also cosmologically acceptable $\relic=0.13 \pm 0.05 $.
Diamonds ($\diamond$) are also cosmologically
acceptable points, but with $\almuon$ within the
region $11 \times 10^{-10}<\almuon<75 \times 10^{-10}$.
Crosses ($\times$) represent the rest of the random sample.
The Higgs boson mass bound $m_h > 113.5 \GeV$ is properly taking
into account.}

\label{fig1} 
\end{figure}

Bearing all these in mind, we proceed discussing our findings.
For our numerical analysis a large random sample of 45,000 points
in the region of the parameter space designed by 
$2<\tan\beta<55$, $M_{1/2}<1.5 \TeV$, 
$m_0 < 1.5 \TeV$, $|A_0|< 1 \TeV$, and $\mu>0$ is used.
The $\mu<0$ case is not favored by the recent
$b \goes s  \gamma$ data, as well as by the observed discrepancy
of the $g_{\mu}-2$, if the latter is attributed to supersymmetry, and  
therefore we shall not discuss it in the sequel.
It is also worth noticing that in the $\mu>0$ case 
the  constraint from $b \goes s  \gamma$ data
is superseded by the $m_h>113.5 \GeV$ bound,
in the bulk of the parameter space \cite{ENO}.
In figure~\ref{fig1} we plot the scalar $\lsp$-nucleon   
 cross section as function of the LSP mass, $\mlsp$.
On the top of the figure the shaded region (in cyan colour) is
excluded by the CDMS experiment \cite{cdms}.
The DAMA sensitivity region (coloured in yellow) is
also plotted \cite{dama}. 
Pluses ($+$) (in blue colour) represent points 
which are both compatible
 with the E821 data $\almuon = (43.0\pm 16.0)\times 10^{-10}$
and the cosmological bounds for the neutralino relic density
$\relic = 0.13 \pm 0.05$. Diamonds ($\diamond$) (in green colour)
are points which are cosmologically acceptable with respect to
the aforesaid bounds, but the bound to the $\almuon$ has
been relaxed to its $2\sigma$ region, namely
 $11 < \almuon \times 10^{10} < 75 $.
The crosses ($\times$) (in red colour) represent the rest of the
points of our random sample.  Here the Higgs boson
mass bound, $m_h > 113.5 \GeV$ has been properly taken into account.
From this figure it is seen that the  
the points which are compatible both the $g_{\mu}-2$ E821 
and the cosmological data (crosses) yield cross sections
of the order of $10^{-8}-10^{-9}$ pb and  the maximum value of  the $\mlsp$
is about $200 \GeV$.  
If one considers the $2 \sigma$ region of the $g_{\mu}-2$ bound the
preferred cross sections can be as small as $10^{-10}$ pb and
correspondingly the upper bound of  $\mlsp$ is drifted up to $350\GeV$. 
In the following figures~\ref{fig2} and \ref{fig3}
the $\sxsec$ is plotted
 as function of the parameters
$m_0$ and $\tan\beta$ respectively. One can see
that the points which  conform  to cosmological and
$1\sigma$ $g_{\mu}-2$ experimental constraints,
yield a maximum value of $m_0$ about $600 \GeV$,
and  for the $2\sigma$ case  $1200 \GeV$.
The aforementioned bounds on the $\mlsp$ and $m_0$
are related with the analogous bounds put on 
the soft  parameter $M_{1/2}$ and $m_0$ 
from the $g_{\mu}-2$ E821 data \cite{ENO,g-2,LS}.

From figure~\ref{fig3} it is apparent  that the majority of 
the points that are compatible with cosmological and $g_{\mu}-2$ data
 are accumulated toward rather large values of  $\tan\beta$,
specifically  $\tan\beta > 40$, although there are indeed few of them
with smaller values of $\tan\beta$.  
As it has been already pointed out in the
large $\tan\beta$ region  we can have simultaneously
cosmologically acceptable values of $\relic$ and also
big values for the elastic cross section $\lsp$-nucleon.
Furthermore the $g_{\mu}-2$ muon data prefer large
values of $\tan\beta$, as $\almuon$ is 
proportional to $\tan\beta$ \cite{LNW}.
Therefore as $\tan\beta$ increases
large regions of the parameter space ($m_0,M_{1/2}$) are
compatible with the E821 experimental constraints.
Taking all these into account it is not surprising that
  the  conjunction of the cosmological
and $g_{\mu}-2$ data happens for large values of the $\tan\beta$ and for large
scalar cross section $\lsp$-nucleon, as it can be perceived from
figure~\ref{fig3}.

Comparing figure~\ref{fig1} and \ref{fig4} one can
realise how $g_{\mu}-2$ data constrain $\mlsp$ mass  to be up to $200 \GeV$ or
$350 \GeV$ for the $1\sigma$ or $2\sigma$ case respectively. 
In figure~\ref{fig4} we don't impose the constraints stemming
from $g_{\mu}-2$ data, therefore  due to the coannihilation processes
the cosmologically acceptable LSP mass can be heavier than
$500 \GeV$.  What is also important   to be noticed
about the direct searches of  DM is that imposing the $g_{\mu}-2$ data  
the lowest allowed $\lsp$-nucleon cross section increased by about
2 orders of magnitude, from $10^{-11}$ pb to $10^{-9}$ pb.
This fact is very encouraging for the future 
DM direct detection experiments.
Figure~\ref{fig5} illustrates the significance of the Higgs boson
mass bound. If one allows for values $m_h>100 \GeV$ many
points which  yielding cross sections even $\mathcal{O}(10^{-7})$ pb
appear. Comparing  figure~\ref{fig1} and \ref{fig5} we observe
that the recent Higgs mass bound ($m_h>113.5 \GeV$) reduces
the maximum value of the scalar cross section for about
one order of magnitude, that is from $10^{-7}$ pb
to  $10^{-8}$ pb. 
There is direct and indirect dependence 
of the $\sxsec$  on $m_h$.
As $m_h$ decreases its contribution to the
$\sxsec$, being proportional to $1/m_h$, increases leading
to larger $\sxsec$.
The indirect relation of the $\sxsec$  to $m_h$ can
be perceived from Eq.~\ref{heavy}. Light $m_h$ results
to light $m_H$ and therefore to large $\sxsec$ again.

Concluding we have  studied the impact of the recent experimental
information to the DM direct searches. Especially  we have
considered the effect of the recently reported deviation
of  $g_{\mu}-2$ from its SM value, as well as
of the light Higgs boson mass bound from LEP experiments.
The imposition of these experimental constraints results
to a maximum value for the spin-independent $\lsp$-nucleon
cross section of the order of $10^{-8}$ pb for
$\mlsp \sim 100 \GeV$ as small as allowed
by chargino searches, 
 and for $\tan\beta > 45$  as large as possible for the  
Higgs states to be as light as allowed by theoretical constraints and 
experimental searches.
As it can be seen from figures~\ref{fig4} and \ref{fig5} 
the effect of these experimental constraints is to
decrease the maximum value of $\sxsec$ by almost
 one order of magnitude, and therefore to make
the direct detection of the LSP on the future experiments
by some means more difficult.

\vspace{3cm}
\noindent 
{\bf Acknowledgements} \\ 
\noindent 
A.B.L. acknowledges support from HPRN-CT-2000-00148
and HPRN-CT-2000-00149 programmes. He also thanks the University of Athens
Research Committee for partially supporting this work. 
D.V.N. acknowledges support by D.O.E. grant DE-FG03-95-ER-40917 and
thanks the CERN Theory Division for hospitality during the 
completion of this work.
V.C.S. acknowledges support  by a Marie Curie Fellowship of the EU
programme IHP under contract HPMFCT-2000-00675.

\clearpage

\clearpage
\begin{figure}[t] 

\centering\includegraphics[height=9cm,width=10cm]{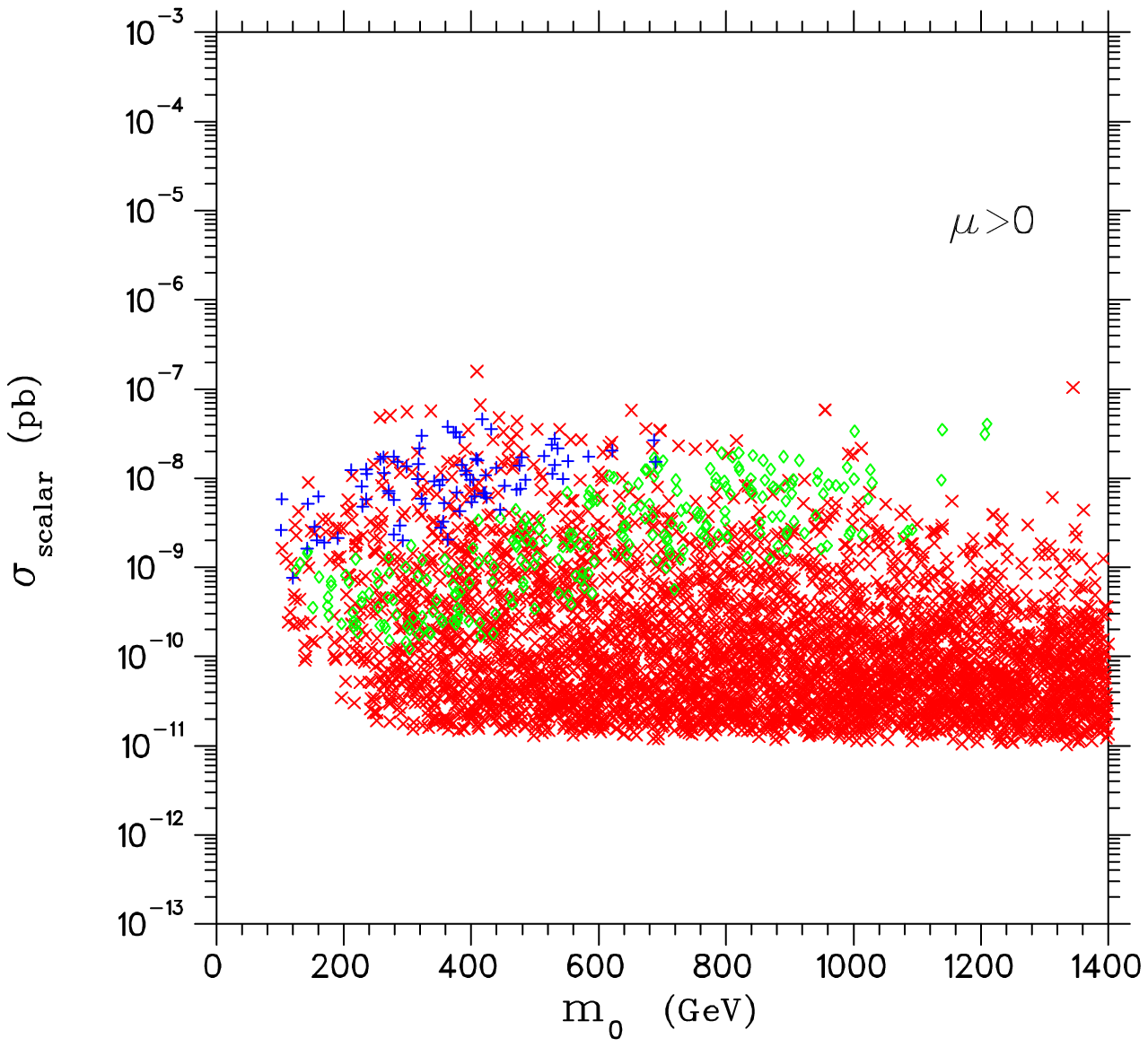}
\caption[]{In this figure we display the
scalar neutralino-nucleon cross section
 versus $m_0$. Points are as in Fig.~\ref{fig1}. } 

\label{fig2} 
\end{figure}

\clearpage
\begin{figure}[t] 

\centering\includegraphics[height=9cm,width=10cm]{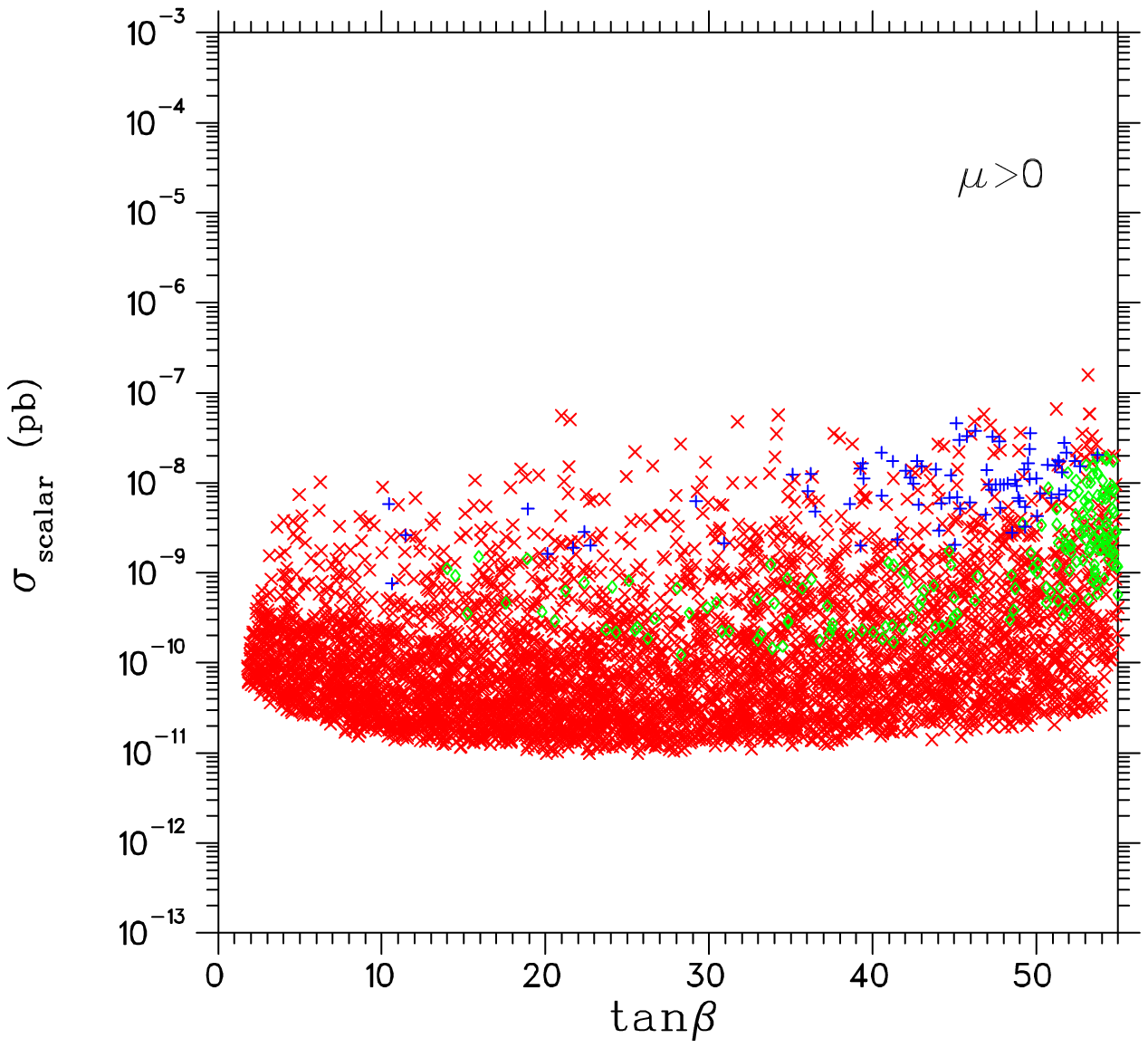}
\caption[]{In this figure we display the
scalar neutralino-nucleon cross section
 versus $\tan\beta$. Points are as in Fig.~\ref{fig1}. } 

\label{fig3} 

\end{figure}

\clearpage
\begin{figure}[t] 

\centering\includegraphics[height=9cm,width=10cm]{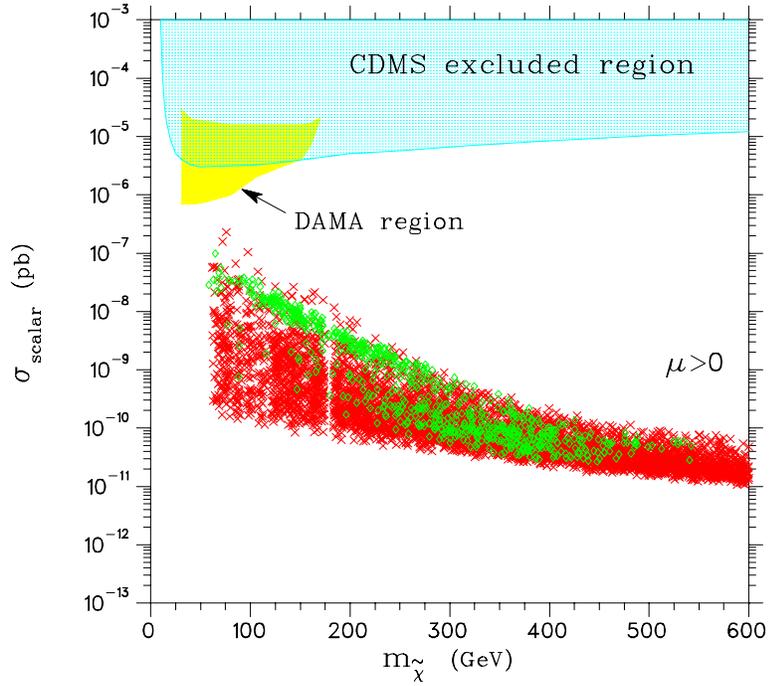}
\caption[]{
Scatter plot of
the scalar neutralino-nucleon cross section
 versus $\mlsp$, from a
random sample of Fig.~\ref{fig1}.  
Diamonds ($\diamond$) are  cosmologically
acceptable points, without putting an restriction from the
$\almuon$. Crosses ($\times$) represent points with
unacceptable $\relic$.}

\label{fig4} 

\end{figure}

\clearpage
\begin{figure}[t] 

\centering\includegraphics[height=9cm,width=10cm]{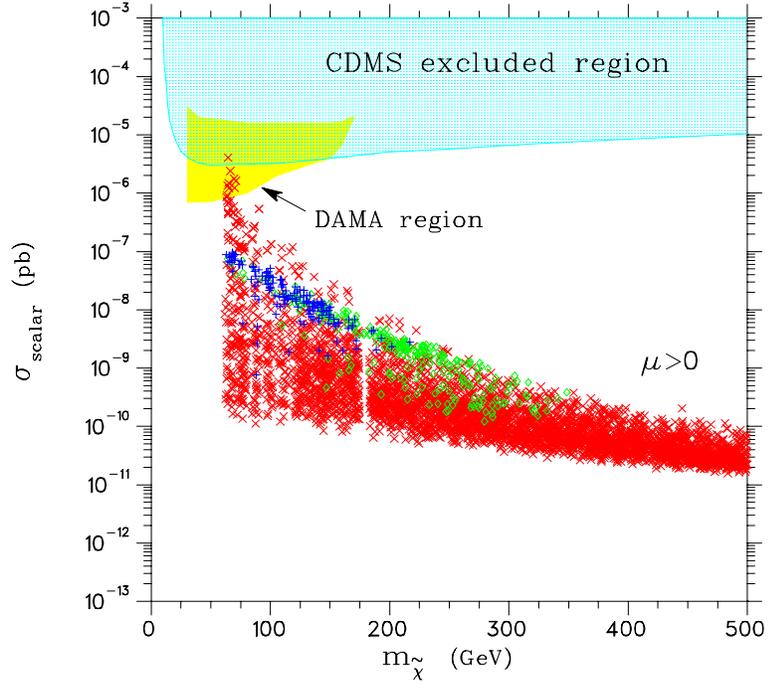}
\caption[]{Scalar neutralino-nucleon cross section
 versus $\mlsp$. Points are as in Fig.~\ref{fig1}.
Here the Higgs boson mass bound ($m_h > 113.5 \GeV$)
has been relaxed and 
the bound $m_h > 100 \GeV $ is used.}

\label{fig5} 

\end{figure}

\end{document}